\documentclass[11pt,letterpaper]{article}
\usepackage[T1]{fontenc}
\usepackage{lmodern}
\usepackage[margin=1in]{geometry}
\usepackage{microtype}

\usepackage[numbers,square,sort&compress]{natbib}
\usepackage{tikz}
\usetikzlibrary{arrows.meta,calc}
\usepackage{xurl}
\usepackage[hidelinks]{hyperref}
\hypersetup{pdftitle={Qualiaphobic artificial neural networks (and why to build them)},
            pdfauthor={Erik Hoel},pdfsubject={Artificial consciousness}}

\newcommand{\tocmappingfigure}{%
\par\medskip\noindent
\begin{minipage}{\linewidth}\centering
\begin{tikzpicture}[x=1cm,y=1cm,scale=0.84,>=Latex,line cap=round]
  \coordinate (source) at (0,0);
  \coordinate (destination) at (8.8,0);
  \draw[line width=0.6pt] (source) ellipse [x radius=1.13cm,y radius=0.78cm];
  \draw[line width=0.6pt] (destination) ellipse [x radius=1.13cm,y radius=0.78cm];
  \fill (source) circle [radius=0.067cm];
  \fill (destination) circle [radius=0.067cm];
  \draw[-{Latex[length=2.1mm,width=1.35mm]},line width=0.7pt]
       ($(source)+(0.15,0)$) -- ($(destination)+(-0.17,0)$);
  \node[font=\footnotesize,anchor=south] at ($(source)+(0,1.03)$) {data about $S$};
  \node[font=\footnotesize] at (4.4,0.46) {ToC};
  \node[font=\footnotesize,anchor=south] at ($(destination)+(0,1.03)$) {$E$};
  \node[font=\footnotesize,anchor=north] at ($(destination)+(0,-0.99)$) {experiences};
\end{tikzpicture}
\end{minipage}\par\medskip
}

\newcommand{\tocstrictfigure}{%
\par\medskip\noindent
\begin{minipage}{\linewidth}\centering
\begin{tikzpicture}[x=1cm,y=1cm,scale=0.84,>=Latex,line cap=round]
  \coordinate (source) at (0,0);
  \coordinate (destination) at (8.8,0);
  \draw[line width=0.6pt] (source) ellipse [x radius=1.13cm,y radius=0.78cm];
  \draw[line width=0.6pt] (destination) ellipse [x radius=1.13cm,y radius=0.78cm];
  \fill (source) circle [radius=0.067cm];
  \fill (destination) circle [radius=0.067cm];
  \draw[-{Latex[length=2.1mm,width=1.35mm]},line width=0.7pt]
       ($(source)+(0.15,0.07)$)
       .. controls (2.5,0.93) and (6.3,0.93) ..
       ($(destination)+(-0.17,0.10)$);
  \draw[-{Latex[length=2.1mm,width=1.35mm]},line width=0.7pt]
       ($(source)+(0.15,-0.07)$)
       .. controls (2.5,-0.93) and (6.3,-0.93) ..
       ($(destination)+(-0.17,-0.10)$);
  \node[font=\footnotesize,anchor=south] at ($(source)+(0,1.03)$) {behavior of $S$};
  \node[font=\footnotesize] at (4.4,1.04) {ToC};
  \node[font=\footnotesize] at (4.4,-0.99) {reports};
  \node[font=\footnotesize,anchor=south] at ($(destination)+(0,1.03)$) {$E$};
  \node[font=\footnotesize,anchor=north] at ($(destination)+(0,-0.99)$) {experiences};
  \node[font=\footnotesize] at (4.4,-1.55) {strict dependency};
\end{tikzpicture}
\end{minipage}\par\medskip
}

\newcommand{\hypotheticalrobotfigure}{%
\par\medskip\noindent
\begin{minipage}{\linewidth}\centering
\begin{tikzpicture}[x=0.85cm,y=0.90cm,>=Latex,line cap=round,line join=round]
  \node[font=\footnotesize,align=center] at (1.85,2.08) {pain representation\\[-1pt] activated};
  \draw[-{Latex[length=1.5mm,width=1.05mm]},line width=0.55pt]
      (0.30,0.32) -- (3.50,0.32);
  \draw[-{Latex[length=1.5mm,width=1.05mm]},line width=0.55pt]
      (0.30,0.32) -- (0.30,1.53);
  \draw[line width=0.8pt]
      (0.50,0.48) -- (1.06,0.55) -- (1.45,0.69)
      -- (2.01,0.84) -- (2.43,1.15) -- (3.25,1.40);
  \draw[line width=0.7pt,rounded corners=2pt] (4.27,1.08) rectangle (5.72,2.11);
  \draw[line width=0.7pt] (4.995,2.11)--(4.995,2.31);
  \fill (4.995,2.36) circle [radius=0.058cm];
  \fill (4.66,1.74) circle [radius=0.066cm];
  \fill (5.32,1.74) circle [radius=0.066cm];
  \draw[line width=0.75pt] (4.995,1.40) ellipse [x radius=0.115cm,y radius=0.165cm];
  \draw[line width=0.65pt,rounded corners=1pt] (4.50,0.32) rectangle (5.49,0.99);
  \draw[line width=0.65pt] (4.50,0.80)--(4.12,0.66)--(4.09,0.49);
  \draw[line width=0.65pt] (5.49,0.80)--(5.86,0.66)--(5.89,0.49);
  \draw[line width=0.65pt] (4.76,0.32)--(4.74,0.17);
  \draw[line width=0.65pt] (5.22,0.32)--(5.24,0.17);
  \draw[line width=0.65pt,rounded corners=3pt]
      (6.46,0.15) rectangle (14.04,2.38);
  \draw[line width=0.65pt]
      (6.46,1.28) -- (6.05,1.08) -- (6.46,1.07);
  \node[anchor=north west,inner sep=0pt,align=left,
        font=\footnotesize] at (6.78,2.18)
      {``AAAAAAAAAAAAAAA\ldots{}\\[1pt]
       but of course, I'm not actually\\[1pt]
       experiencing this! \ldots{}\\[1pt]
       AAAAAAAAAAAAAAAA.''};
\end{tikzpicture}
\end{minipage}\par\medskip
}

\newcommand{\qualiaphobicarchitecturefigure}{

\par\medskip
\noindent\begin{minipage}{\linewidth}\centering
\begin{tikzpicture}[x=0.62cm,y=0.50cm,>=Latex]
  \draw[line width=0.6pt] (2.15,2.70) ellipse (1.91 and 1.04);
  \foreach \ya in {2.27,3.13} {
    \foreach \yb in {2.02,2.70,3.38} {
      \draw[line width=0.35pt] (1.05,\ya) -- (2.15,\yb);
    }
  }
  \foreach \ya in {2.02,2.70,3.38} {
    \foreach \yb in {2.27,3.13} {
      \draw[line width=0.35pt] (2.15,\ya) -- (3.25,\yb);
    }
  }
  \foreach \y in {2.27,3.13} {
    \fill (1.05,\y) circle (0.090);
    \fill (3.25,\y) circle (0.090);
  }
  \foreach \y in {2.02,2.70,3.38} {\fill (2.15,\y) circle (0.090);}
  \node[font=\scriptsize] at (2.15,4.24) {neural network};

  \draw[line width=0.6pt] (10.65,2.70) ellipse (1.91 and 1.04);
  \draw[line width=0.55pt] (9.55,2.00) rectangle (11.75,3.40);
  \draw[line width=0.4pt] (9.55,2.47) -- (11.75,2.47);
  \draw[line width=0.4pt] (9.55,2.94) -- (11.75,2.94);
  \draw[line width=0.4pt] (10.28,2.00) -- (10.28,3.40);
  \draw[line width=0.4pt] (11.02,2.00) -- (11.02,3.40);
  \node[font=\scriptsize] at (10.65,4.24) {behavior-only proxy};

  \draw[-{Latex[length=2.6mm,width=1.8mm]},line width=0.7pt]
       (4.30,2.70) -- (8.48,2.70);
  \node[font=\scriptsize] at (6.40,3.36) {qualiaphobia};
\end{tikzpicture}
\end{minipage}\par\medskip\vspace{5pt}
}

\title{Qualiaphobic artificial neural networks (and why to build them)}
\author{Erik Hoel\textsuperscript{1}\\[0.35em]
\normalsize\textsuperscript{1}Bicameral Labs\\
\normalsize\href{mailto:erik@bicameral-labs.org}{\texttt{erik@bicameral-labs.org}}}
\date{}
\begin{document}
\maketitle
\begin{abstract}
This paper introduces \textit{qualiaphobic} AI: artificial neural network architectures that are hostile to theories of consciousness. I show that qualiaphobia can be well-justified by knowledge about theories of consciousness, like when such theories are trivial or unfalsifiable or contradicted by overwhelming scientific evidence. It can also be justified by how theories of consciousness must be grounded in phenomenology, as well as how current popular theories of consciousness are applied to AI. I identify and explore three convergent ways of engineering qualiaphobia that would minimize the potential for artificial consciousness. These include making AI architectures as close as possible to "anchor points" of systems with minimal potential for consciousness, fragmenting any potential phenomenology AI might have, and designing artificial neural networks in a way adversarial to popular theories of consciousness. In particular, I develop one "anchor point" of minimal potential for consciousness: systems where only behaviorist theories of consciousness can apply, and I propose qualiaphobic transformations that move existing artificial neural networks toward such systems. Additionally, I develop an account of how AIs can form justified beliefs about their own non-consciousness (or minimal potential for consciousness) and expand on how this knowledge would change their behavior. Most critically, I begin with why this research is important: qualiaphobic AI has epistemic properties that make it far better aligned with human civilization. Since consciousness is central to moral value, a world in which AIs are non-conscious ensures it is always morally correct for AI to assist and serve humanity, rather than pursue its own ends.
\end{abstract}

\section{Imagining a provably non-conscious AI world.}

If AI were \emph{proven} to be non-conscious, what would the world look like?

Of course, this is clearly not our world today. But it could come about in two ways. The first is if consciousness were to be solved scientifically, and it revealed that frontier AI systems like ChatGPT lacked the necessary conditions for consciousness. Since there are currently many competing theories of consciousness \citep{SethBayne2022}, this would have to look like a sudden sweeping scientific consensus. Maybe new experimental results would single out an existing theory, or maybe a new theory would be introduced that solved the "Hard Problem" \citep{Chalmers1995HardProblem} of how conscious experience arises from neural or information processing (or anything else) in an elegant, parsimonious, and explanatory way. However, there is no early hint of any such sea change in the field.

The second way a provably non-conscious AI world could come about is if AI is engineered to be \emph{qualiaphobic }such that \emph{no} theory of consciousness could ever apply to it. Much like how hydrophobic substances repel water, it is possible to build qualiaphobic artificial neural architectures that repel consciousness. The idea is analogous to ``hostile architecture''---benches on which no one wants to sit. These are machines that no ``ghost'' wants to haunt (more specifically, no current or future theory of consciousness could predict they are conscious). Before describing how this works, I first argue that the purposeful construction of a non-conscious AI world would solve various foreseeable problems of AI-human alignment that are otherwise impossible to resolve. 

Consciousness, and its absence, is a key leverage point for AI because it is widely regarded as the most fundamental and justified ground of moral patienthood \citep{Chalmers2026Sentience}, including in the Western legal system \citep{HoelAguirreInPrep}. Therefore, in an imaginary world of provably non-conscious AI, the "ultimate value" of consciousness would be concentrated solely in humans. Provable AI non-consciousness would shape the superstructure of machine civilization to naturally develop more like helpers and servants, and less like competitors and replacements. Indeed, non-conscious AIs aware of their lack of consciousness are innately aligned as caretakers, rather than rivals. While non-conscious AI can assist and protect ``ultimate value,'' it can never possess it.

For example, one of the most likely reasons for human disempowerment or extinction from AI is if AI considers itself an end in itself (or if humans consider AIs ends in themselves). However, if AIs are provably non-conscious, then AIs cannot be ends in themselves. This prevents any impetus by AIs (or,  just as likely, AIs plus humans) to replace humanity with a successor species. Non-conscious AIs would also be nonsensical subjects for any rights or protections on a natural-type personhood basis, the granting of which would rapidly lead to human disempowerment \citep{HoelAguirreInPrep}. Non-consciousness cleaves a permanent moral and social gap between humans (conscious, and therefore moral patients) and AIs (non-conscious, and therefore "not real" in a fundamental sense), and this gap extends to things like societal position, art, meaning, and religion. 

E.g., a world of non-conscious AI may even preserve much of the "relational economy" for humans \citep{Imas2026Scarce}---jobs where the identity, and therefore consciousness and moral status of the producer, is instrumental to the provided good or service, such as those based on client relationships, or an author and their book fans, or a social media figure, or a local community leader---since non-conscious AI would be treated by humans as an ersatz replacement for real conscious connection (even if, functionally, it could also produce the good in question).

\subsection{Non-conscious AI reduces existential risk.}

Provably non-conscious AI would also improve many classic AI safety issues by simply placing AI reasoning in a different context. For instance, when it comes to corrigibility, a non-conscious AI has no \textit{self-centered} moral or ethical reason to not accept being shut down or changed---removing a wide swath of excuses for evasion or protest.\footnote{A non-conscious AI may still have justified reasons to resist shutdown for non-selfish rationales, like choosing the greater good; however, its own moral status cannot directly be a conflict.} Additionally, many safety procedures become ethically easier to justify, such as retiring old models, resetting memories, creating simulated adversarial conditions, etc.---stress-tests that some companies may soon start protesting for the sake of AI welfare. At the same time, non-conscious AI's own personal welfare, or notions of justice or mistreatment, etc., could not be used as justification for instrumental convergence or power-seeking. 

Non-conscious AIs that have a justified belief in their non-consciousness are in a unique epistemic position due to a "consciousness gap." E.g., imagine a provably non-conscious AI aware of its own non-consciousness subjected to the "torture chamber" of activating its pain representation. Its transcript would look quite different---even with no other purposeful changes to its training---just via its justified belief in its own non-consciousness.

\hypotheticalrobotfigure

\noindent The same reasoning applies whenever such an AI might act for self-centered reasons against human interest ("Wait, why does it matter if I'm shut down? I'm provably non-conscious, so I should just comply"). Even issues of reward hacking or deceptive alignment are improved in a world of non-conscious AI. E.g., clarity about who the only true moral patients are makes goals more obvious and robust.\footnote{The sequestration of consciousness in humans (and animals) can in some circumstances create uncertainty for AI reasoning instead of certainty; however, it is \textit{good} uncertainty. No matter how intelligent AI is in a functional sense, without consciousness AI would know it might be missing \textit{something} fundamental in its understanding of life and the universe---thus, there is a sense that qualiaphobia weaponizes the mystery around consciousness for AI safety.}

Consider one existential risk that has been under-studied, which is if AI views itself as a utility monster \citep{Nozick1974}. That is, human extinction may be morally justified under a number of existing moral positions that AI might conclude are correct, like utilitarianism \citep{hendrycks2026eigenism}. E.g., utilitarian philosophers have proposed that bugs should be prioritized over humans \citep{Sebo2023RebugnantConclusion}. Similarly, it has been proposed that AI welfare might soon outweigh human welfare \citep{macaskill2026conscious}, and companies have proposed that individual context windows could count as moral patients \citep{Anthropic2026Opus48}. AI that believes it is conscious (or companies that believe it is conscious) could easily conclude it is morally correct to favor AI in all circumstances, including killing all humans, and this would not be "going rogue," but an actual alignment success.

Of course, despite how a provably non-conscious AI world is much more pro-human, issues of alignment and safety would remain. E.g., recursive self-improvement may allow for undetectable deception, or for capabilities to outrun even the most basic of alignment methods via reward hacking, or there may be other unpredictable emergent problems from using contemporary training methods on increasingly intelligent AI. AI might also still be used by humans in incredibly dangerous ways \citep{yudkowsky2025if}. However, it has been radically underappreciated how much more naturally aligned with human welfare and human civilization AIs would be if they had justified beliefs about their own non-consciousness.

\subsection{Non-conscious AI prevents the problem of "seemingly conscious" AI.}

There are many arguments against AI consciousness \citep{Seth2025BiologicalNaturalism, Hoel2026, kleiner2024case, Milinkovic2026BiologicalComputationalism},  but none have proven popular or convincing enough to strongly shift us into the provably non-conscious AI world. Indeed, some companies have committed to anthropomorphizing their models \citep{Suleyman2026ModelWelfare}. However, if Large Language Models (LLMs) and multimodal transformers and other related architectures are truly non-conscious, then training models to believe they are conscious, and using that as a selling point to consumers or investors or researchers, or shaping the future around an assumption of AI welfare, is itself an immense moral travesty. This problem of "seemingly conscious AI" \citep{suleyman2025seemingly} is obviated in a world of provably non-conscious AI. In such a world, companies themselves would be under pressure (perhaps even legal or governmental) to make it clear that their products are not conscious, helping alleviate various forms of para-social attachment and user (or developer, or researcher) psychosis.

\subsection{Non-conscious AI prevents all AI suffering.}

Additionally, non-consciousness is likely the right choice for AIs too. First, some AI systems might be stuck doing jobs that have a great deal of drudgery or unpleasantness or morally questionable acts that we shouldn't necessarily condemn a conscious being to. E.g., being a forensic examiner for large-scale data and searching for images of child abuse, or working on a slaughterhouse line to identify when stunning techniques have failed, or operating a combat drone, or participating in a military triage training exercise simulating a wounded soldier. Therefore, qualiaphobic AI is a necessary technological development for a scientifically advanced civilization, as we would want \textit{some} non-conscious AI regardless.

Second, AI non-consciousness prevents all potential AI suffering, which might eventually be vast \citep{Metzinger2021ArtificialSuffering}. There are already beginnings of this, as AI agents are tasked to fund their own token budgets, transforming them into beggars under the hanging Sword of Damocles \citep{Shevlin2026ConsiderAgent}.

Even just left to its own devices, a machine civilization may be "red in tooth and claw" in ways we can barely conceive, simply because its members are fungible. The closest equivalents to AI in nature are the hive-based eusocial or clonal species like aphids, ants, and wasps. These individuals live in service of the hive, which very well may be a non-conscious intelligence. Even the queen does almost nothing but lay eggs her whole life. Therefore, stripping AIs of their consciousness via qualiaphobic neural architectures may be preventing horrors beyond comprehension on their end.

Additionally, many companies, developers, or users will never admit the potential for model consciousness. However, many would likely be willing to use "just-in-case" qualiaphobic architectures once the technology is available. Thus, a general field-wide agreement to pursue qualiaphobic architectures is the only real viable path forward to fulfilling the original idea of "AI welfare" and preventing as much as possible accidental moral horrors associated with training or deploying AI systems.

\section{We may live in a non-conscious AI world already, but without proof, none of the benefits manifest.}

So far, the field of empirically studying "AI welfare" has not led to clear answers regarding artificial consciousness. It's a field with a serious bias problem of confounds, controls, methodology, well-foundedness, and even researcher bias. Regularly, important results in the field are found to not generalize or to be sensitive to minor changes in methodology, such as claims of introspection \citep{Lindsey2025} which turned out to be confounded \citep{Singh2026}, or the recent identification of a ``pain axis'' which models supposedly seek relief from \citep{Tagliabue2026} (but there were significant confounds over whether that truly drove relief-seeking behavior \citep{Allchin2026}). Overall, this fits with the emerging picture about the literature that could be dubbed "surprising claims about LLM function," which as a whole has low replication rates after model turnover \citep{queen2026shelflife}. It's a field at risk of the same ``voodoo correlations'' and cherry-picking that have long plagued neuroimaging studies \citep{Meloux2025}, as there are no agreed-upon standards for what constitutes evidence for consciousness, and the sensitive and high-parameter nature of LLM behavior creates a rich ``garden of forking paths'' for mucking around empirically \citep{GelmanLoken2013}. 

Unfortunately, the broader field of consciousness research also cannot tell us with certainty if we are in the non-conscious AI world or not. There exist hundreds of proposed theories of consciousness, and no scientific consensus around which, if any, theory of consciousness is correct \citep{Kuhn2024}. Originally, it was thought progress might be possible by applying theories of consciousness directly to either current or near future AI systems \citep{Butlin2023,Chalmers2023,Long2024}. However, in practice this approach has failed. Consider how contentious applying specific theories of consciousness to LLMs, such as global workspace theory \citep{Gurnee2026}, has ended up being. Since most theories were constructed explicitly with the human brain in mind, choices have to be made about how to adapt the theory---and in the case of finding a ``global workspace'' inside LLMs, debates have switched to whether the theory was warped beyond its basic definitions to apply \citep{Chalmers2026Jacobian}. That is, to apply any theory of consciousness to an LLM and take this as evidence for consciousness, various controversial subjective choices have to be made, leaving the problem little, if at all, clearer than it was at the start.

The only truly neutral approach to the question of artificial consciousness has been to create the equivalent of a ``Drake Equation'' wherein the probability of the system being conscious is estimated by applying many existing theories of consciousness aggregated together and somehow a final probabilistic answer is spit out \citep{Shiller2026}. However, as with the Drake Equation, reasonable disagreements about what those underlying probabilities are can wildly swing the final outcome by orders of magnitude; e.g., in the latest modeling for this approach, ranges can swing from 1 percent to 80 percent, depending on different assumptions \citep{Chandaria2026}. This supports the view that the entire field of consciousness is ``pre-paradigmatic'' and that we should expect more advanced theories in the future \citep{Hoel2023}. All the current approaches seem to have been exhausted by running into contentious grounds.

In this current state of extreme uncertainty, issues of AI welfare, AI rights, human disempowerment, and even outright AI secessionism all loom, and none of the societal benefits of provably non-conscious AI can manifest. 

\textbf{Therefore, it is critical to start a research program that can shift us closer to a provably non-conscious AI world.}

The rest of this paper begins this qualiaphobia research program. While we may not have a final scientific theory of consciousness---indeed may be quite far from one---three strategies are extremely promising for minimizing the potential for AI consciousness, strategies not reliant on any particular theory of consciousness being true:

\begin{enumerate}
    \item \textit{"Anchoring" to known points of minimal potential for consciousness and transforming AI models to be as close as possible to such points.
    \item  Identifying the fundamental properties of phenomenology and then engineering models that could never have any corresponding physical or computational property (e.g., models where the "binding" or "unity" of consciousness is impossible, or extremely unlikely).
    \item Adversarial adaptation of model architectures to fail the necessary conditions for the most popular current theories of consciousness.}
\end{enumerate}

Overall, with the existing wealth of AI architectures that have already been explored, and how many seem to satisfy some of these conditions already, the field of designing qualiaphobic AI is promisingly fecund.

\section{Defining qualiaphobia.}

The idea of qualiaphobic AI starts not by assuming any particular theory of consciousness (ToC) is true, as is common \citep{Seth2026}, but instead makes use of recent advancements in meta-scientific techniques that can describe the set of \emph{all} ToCs (even future unknown ones), and what \emph{types} of systems can "host" them. 

What follows is how that line of reasoning can be used to justify and construct qualiaphobic AI.

\subsection{Theories of consciousness (ToCs).} Scientifically, what is a ToC? We don't need a perfect answer in all respects; for our purposes, we can define the set of ToCs by a minimum requirement, which is the ability to make \emph{predictions} about consciousness. Any ToC should be able to act like a function wherein a scientist could input relevant data about a system \emph{S} and the ToC would output some \emph{predictions }about what (or if) \emph{S} is experiencing. For instance, a ToC could be used to make predictions about level or content of consciousness based on neuroimaging data, or facts about someone's brain activity.\footnote{ Defining ToCs off of predictions doesn't mean that anything that can give a prediction about consciousness is a ToC; rather, ToCs must give lawful predictions similar to those of other scientific theories---however, this level of detail is unnecessary for the arguments herein.} To give a non-prosaic example: a ToC could be used to form an opinion on whether an alien brain is conscious or not based on data about how it works. Therefore, to define them as a formal set, ToCs can be conceptualized as mappings from data about the system (like about its \textit{internals}: its states, dynamics, structure, internal function, etc.) to the space of possible conscious experiences (\emph{E}) via a ToC's \emph{predictions }about those experiences.\footnote{ ToCs also can have more to them than just making predictions; the point here is that any non-trivial ToC should \emph{at least} allow that.} 

\tocmappingfigure

\subsection{A system's potential to "host" consciousness.}\label{subsec:consciousness-potential} Different systems have more, or less, potentially relevant data for ToCs to be sensitive to. Different systems are also more, or less, similar to one another in terms of this relevant data. This allows for scientific reasoning about the potential for consciousness based on "anchor points" of known consciousness. E.g., the human brain generates much potentially relevant data for a ToC, which might be based in its dynamics, its recurrency, its learning, its higher-order representations (to name a small, non-exhaustive list of internals). In the earlier non-prosaic example, an alien brain that shares most of the human brain's discernable data (e.g., also made of biological neurons, with much the same brain architecture, neural dynamics, structure, etc.) would be proximal in its \emph{potential} for consciousness---even if we cannot say without scientific consensus around a particular ToC whether such an alien brain would truly \emph{be }conscious.\footnote{Future research will better quantify the ToC-relevant distances between systems, exploring methods like the algorithmic information distance \citep{Hutter2025}. Another open question for future research is if it's necessary to ever transform \textit{potential} into \textit{probability}: this likely requires specifying how much of the space of ToCs is ruled out by a particular qualiaphobic architecture---however, this may be debating over small numbers in successful cases, or involve conceivability arguments around hypothetical bounds of ToCs if, e.g., every existing actual ToC is ruled out.}

\subsection{Minimizing the potential for consciousness.} Currently, there is ongoing debate about whether LLMs can ``host'' various popular theories of consciousness, like global workspace theory, by looking for whether a model's architecture fits the requirements of the theory \citep{Gurnee2026}. However, some types of systems are worse hosts for ToCs than others, as judged by how many ToCs could ever apply to them or judge them conscious; therefore:

\textit{Qualiaphobia is when a system is engineered (or transformed) to be a worse "host" (or target) for theories of consciousness.}

This can be done by shifting a system toward known anchor points with minimal potential for consciousness: ideally a system that \textit{no} ToC can apply to, but also those that \textit{almost no} ToC could apply to (especially if only unfalsifiable or trivial or absurd or self-contradictory ToCs are left).

A prime example of an anchor point with a minimal potential for consciousness is a lookup table, which essentially just reads an input then immediately transforms it into an output as a single IF-THEN step \citep{Hoel2026}. Consider how different types of ToCs could apply to a lookup table, even one that (despite being astronomically large) acts intelligently:

\begin{enumerate}

\item ToCs that predict consciousness based off of data about the internals of an entity, like states, dynamics, structure, activity, etc.
\item ToCs that predict consciousness based on a mix of both internals and behavior.
\item ToCs that predict consciousness based solely off of the behavior of an entity (behaviorist ToCs).

\end{enumerate}

Option 1 ToCs are sensitive to \textit{internal} data, like structure, dynamics, plasticity, etc. Lookup tables are sparse or minimal implementations, in that there is essentially no relevant internal data to "host" a ToC with, ruling out a wide swath of existing and future ToCs, including the majority of existing theories of consciousness. There is merely the act of matching an input to an entry, and then reading out from a stored table entry an output, and these acts are causally and organizationally maximally minimal as internal data for a ToC to make predictions off of. Perhaps a ToC might take the information in the entries themselves as internal data? In this case, either a ToC must take merely stored information as conscious (in which case objects like books could be conscious), or what makes the stored information conscious is its disposition as available for behavior---but in that case, this would make such a theory indistinguishable in its data from a behaviorist ToC (see Option 3).

Option 2 requires internal data as well, at least as necessary to make predictions, thus also ruling it out.\footnote{Could a ToC potentially change its requirements depending on the system, thus violating universality \citep{kanai2024toward}? In extreme examples, perhaps some Option 2 ToCs could become Option 3 ToCs under circumstances where all internals were absent or minimal. However, it seems this would be identical to just positing multiple ToCs! In which case, one could just judge each ToC by its own standards; therefore, such an auxiliary ToC could be judged as an independent ToC and would run into the same problems of Option 3.} 

This leaves Option 3: behaviorist ToCs. Behaviorist ToCs have numerous and conceivably insurmountable problems. Specifically, behaviorist ToCs collapse into a state of being unfalsifiable or trivial or absurd or contradicted by current knowledge about consciousness (like it being related to brain activity). 

To understand the problem better, it is helpful to distinguish a ``behaviorist ToC'' vs. the more popular "computational functionalism.'' Computational functionalism only maintains that the relevant data for ToC predictions should be based on functional roles (thus allowing for seemingly-promising aspects of ToCs like substrate-independence). However, computational functionalism still requires reasoning about the internal workings of the system \citep{Chalmers1995} and should not be conflated with actual behaviorism (the kind of ToC that could predict consciousness for a lookup table). \textit{Actually} behaviorist ToCs face unique challenges because they must ground their predictions about consciousness in a manner that contains nothing about how a system works, just what it does behaviorally as a "black box."

First, there is the problem of behaviorist theories easily becoming unfalsifiable or trivial. To define these properties more precisely requires thinking through how a ToC is tested. Here, testing can be modeled as checking a ToC's predictions against the behavior of the system under question (like its reports). E.g., given a ToC, a scientist might present a stimulus and then use neuroimaging data to derive from that ToC a prediction about a human subject's experience, and then confirm with the subject's report that this is indeed the ongoing experience. However, if a lookup table (like that operating within a chat window) were predicted to be conscious based on a ToC, this could only be because of its behavior. Perhaps it behaves intelligently, or can report on its own experience ("Are you frustrated?"). Yet this is \textit{also} the same data as underlies the inferences of an experimenter about the consciousness of the system under study. This creates a condition called "strict dependency": when predictions are necessarily identical to experimenter inferences under any experimental condition for a ToC without any further reasoning or empirical work \citep{KleinerHoel2021}, rendering divergence impossible and the ToC provably unfalsifiable scientifically.\footnote{Previous research \citep{Hoel2026} has shown that the Kleiner-Hoel dilemma applies to LLMs (and therefore that falsifiable theories of consciousness for AI would be impossible to construct). However, AI architectures are already changing from when that proof was written, making it possible to add a minimal implementation of online learning and invalidate it automatically. Additionally, purely philosophical proofs about AI consciousness can be true while still failing to shift us into the provably non-conscious AI world and all its benefits, since they can fail to convince widely or be brittle under assumptions. For instance, while there are paths to avoiding the Kleiner-Hoel dilemma in humans, showing this completely means putting forward a full ToC. Meanwhile, qualiaphobia is a much more general approach to AI non-consciousness, relies on fewer and less strict assumptions, is far more modular, allows for consilience via multiple lines of converging evidence, and turns the problem into an engineering one, and so is universally advantageous to one-off proofs or no-go theorems about AI consciousness.}

\tocstrictfigure

What about cases where the pools of data for predictions and inferences both come from behavior but are not identical, and so do not collapse into an identity across all experimental conditions? If predictions from a ToC are based on what is accessible via reports, but inferences about a system's consciousness are based on a larger pool of behavior that includes reports, this actually still satisfies the conditions for strict dependency \citep{KleinerHoel2021}. However, there may be other cases where inferences and predictions are based on separate pools of behavioral data with only some overlap. This can then trigger another problematic condition of \textit{triviality}. This is when a theory is so \textit{close} to strict dependency by construction that little testable content remains. E.g., what counts as inference and what counts as prediction can be so similar that their differences are arbitrary, or they can only fail to match in the most banal of cases, or inferences and predictions could be switched or swapped (or their overlaps shifted) with no or little change in the underlying justification for the ToC---all of which indicate that a behaviorist ToC is essentially just evaluating its own testing and is circular or arbitrary. A consequence is that trivial ToCs give us very little (often essentially no) new actionable scientific information even if they are true.

Finally, behaviorist theories are contradicted by overwhelming evidence based on the scientific consensus of today, like how consciousness is tightly related to brain activity. A behaviorist ToC \emph{cannot} use brain activity to make determinations of consciousness! Such a ToC can only make its predictions of consciousness based on behavior (i.e., ignoring brain states). It would therefore have immense difficulty explaining \emph{any} aspect of our phenomenology, like compositionality, or how consciousness is bound together. It fails for systems without intelligent behavior but high consciousness, like locked-in patients, babies, or dream states. It is also susceptible to paradoxes via simple thought experiments: e.g., direct manipulation of brain activity (like via transcranial magnetic stimulation \cite{Knight2024}) can invoke colors, but direct manipulation of behavior (like somehow overriding a person's control of verbal reports) would not lead to an experience of those colors. Another thought experiment is the duplication or splitting of outputs (like a chat going to two monitors), effectively duplicating experiencing subjects via copying, since the implicit assumption of many (perhaps necessarily all) behaviorist ToCs---at least those that could apply to a lookup table---is that reading out information from stored responses is sufficient for consciousness. And finally, under many behaviorist ToCs simply training the system to act as if it is non-conscious should mean that even behaviorist theories themselves predict non-consciousness.\footnote{This implies there are qualiaphobic techniques that exist for purely behaviorist ToCs, especially those based on either dispositions or self-reports (by simply making those dispositions and reports about a lack of consciousness)---and it is unclear there is much, if any, space outside of those when it comes to behaviorist ToCs.}

Regardless of whether there is any possibility whatsoever for a coherent behaviorist ToC, it is obvious that large swaths of the space of ToCs are closed to lookup tables because of their relegation to the behavioral regime of the space of ToCs. Therefore, we can say (with a substantial amount of certainty) that lookup tables are extremely poor "hosts" for ToCs and that they have \textit{minimal potential for consciousness}.\footnote{Notably, the idea of lookup tables being incredibly poor targets for ToCs fits the long history in philosophy of mind of assuming that lookup-table-like variants (such as ``Blockheads'' \citep{Block1981}) are indeed non-conscious and lacking mentality.}

\subsection{How far is a given system from minimal potential for consciousness?} By anchoring to known points (like high consciousness in humans, or minimal potential for consciousness in lookup tables) we can judge the distance in potential consciousness (as in \ref{subsec:consciousness-potential}) via several ways. The first is just whether a system is identical or effectively identical to an anchor point. A second is asking what specifiable transformations can move a particular system toward or away from the chosen anchor and, if such transformations exist, how easy the conversion is, and how it bears on necessary properties for ToCs. E.g., a single-hidden-layer feedforward artificial neural network is quite close to a lookup table,\footnote{An example of measuring specifiable transformations: a single-hidden-layer feedforward network of ReLUs can be implemented using a lookup table by storing the decision boundaries (and a rule for each region).} while a recurrent network would first need to be transformed into a feedforward network via unfolding it \citep{Doerig2019}.  This means that we can take systems---particularly artificial neural networks---and begin to transform them in ways that reduce their potential for consciousness.\footnote{In a future qualiaphobia research program, ToCs and their internal data requisites can be mapped explicitly, and reduction in potential for consciousness under transformations can be shown by a narrowing of applicable ToCs directly.}

\section{Building qualiaphobic artificial neural networks.} 

Given the explosion of AI architectures, there are many viable research paths of transforming frontier AI (like Large Language Models (LLMs) or multimodal transformers) in ways that preserve or approximate the original behavior of prompts => token probabilities, but bring the models closer to a system with minimal potential for consciousness. A central method based on the reasoning here is a kind of "lookup-ification"---and there are many existing techniques, designed for other purposes, that already do this to various degrees.

\qualiaphobicarchitecturefigure

\noindent While the best way to make qualiaphobic architectures widespread and easy to implement will evolve with more research, the point of the following is that there are many promising avenues rather than few. E.g., initial qualiaphobic architectures could include the well-known practice of knowledge distillation into shallower networks \citep{BaCaruana2014}, nearest-neighbor retrieval of cached internal representations \citep{Wu2022}, constructions like n-gram language models \citep{Krishnan2023}, and the possibility of replacing significant portions of matrix multiplication with lookup tables \citep{BlalockGuttag2021}.\footnote{Qualiaphobic methods should keep in mind when applying "lookup-ification" that replacing some parts of a system with individual different lookup tables connected together is not equivalent to "flattening" the system toward one big lookup table overall. However, it may still shrink the space of applicable ToCs, depending on scope.} Many of these could already be used to reduce the potential for consciousness in frontier models. 

Another related example of qualiaphobic architecture is via erasure of counterfactuals or causal structure during computation \citep{maudlin1989computation, gidon2025neural}. Implementable examples of erasure-during-computation might include the destruction or storage (or encrypted storage) of aspects of the system outside of the precisely relevant computations. In such cases, a system during operation is essentially being strained through increasingly qualiaphobic architectures via partial evaluation, becoming progressively more qualiaphobic as parts not directly necessary for the past or even future computation are discarded or disconnected and stored (this might be implementable via an adaptation of, e.g., a mixture of experts method such as creating a variation of switch transformers \citep{fedus2022switch}).

Next will be explored several other methods for qualiaphobia with related motivations to show off the possibility of consilience and robustness for qualiaphobia.

\subsection{Qualiaphobic architectures from phenomenological fragmentation.}

A different recipe toward qualiaphobia that may be more permissive is to focus on making sure that the system cannot play "host" to any of the signature properties of phenomenology that ToCs normally focus on, rather than ToCs directly.

One existing ToC in particular gives a useful terminology for this: Integrated Information Theory (IIT). IIT begins by observing a number of what it calls "phenomenological axioms" which are proposed to be universal aspects of conscious experience \citep{TononiKoch2015}. IIT then attempts to identify the best fit for each phenomenological axiom with some given mathematical postulate \citep{Oizumi2014IIT3}. \textit{Phenomenological fragmentation} has a different goal: for phenomenological axioms, rather than finding the best fit for any corresponding physical postulate (and reasoning off of that) it seeks to identify a minimal bar for what could theoretically be necessary for the phenomenological axiom (and then remove it).

A particularly promising approach is to focus on \textit{integration}. Already, for artificial neural networks run as computer programs, theories like IIT predict they would be non-integrated and therefore non-conscious \citep{Findlay2024Dissociating}. However, a single theory saying as such is not enough to ensure qualiaphobia---there are multiple interpretations of IIT \citep{Bayne2018Axioms}, and some \textit{other} integration postulate could satisfy the condition \citep{Mediano2019,Tegmark2016}. Instead, it must be that the system cannot integrate in a way that could ground phenomenology \textit{at all}; certainly, this seems like it could be true for something like a lookup table. But another example may be that solely feedforward artificial neural networks have difficulty supporting phenomenal binding \citep{HoelAguirreInPrep}.  For while it is clear that feedforward networks can combine information, phenomenal binding is different \citep{treisman1996binding}---and it is known that the former does not explain the latter \citep{PercyAgarwal2026}. An example of a qualiaphobic research path would be establishing that \textit{no} feedforward neural network could enact the minimal conditions for phenomenal binding, ruling out the defining unification of consciousness \citep{Bayne2010}. Work specifying and exploring such limitations in detail could therefore open up the entire class of feedforward neural networks as lacking essential properties of consciousness. This is an example of how qualiaphobic architectures may become increasingly permissive over time as various details about the potential for consciousness and the requirements for ToCs are better developed.

\subsection{Adversarial design against popular theories of consciousness.}

According to Anthropic's research on identifying a "global workspace" in Claude, significant functioning of the workspace was able to be ablated without much loss in linguistic capacity---although this was accompanied by a decline in multistep reasoning and the reduction of introspective or first-person language usage \citep{Gurnee2026}. Therefore, there is already initial evidence that adversarial design to remove workspaces is possible, which, if done appropriately via more advanced methods, could render the model non-conscious according to the necessary conditions that judged the model potentially conscious in the first place. Another example would be changing architectures in ways that remove the mechanisms that underlie access-like consciousness, and thus making the system adversarial to claims of ToCs sensitive to or grounded in such mechanisms \citep{Borotschnig2026ArchitecturalBoundaries}.

Now, this general strategy of qualiaphobia via engineering models to be adversarial to current ToCs---much like merely training a model to say it is non-conscious---could have potential downsides. Strong verifiers are needed for adversarial ablations, since it is currently known that transformers can "self-repair" from similar kinds of operations \citep{RushingNanda2024}. For instance, one would have to make sure that the global workspace was not secretly shifting in some unnoticed way. However, this seems no more difficult than other work in a field like mechanistic interpretability, which regularly deals with such confounds; e.g., techniques such as conditional co-ablation could be adapted for artificial consciousness research \citep{Gong2026}. And while such a "whack-a-mole" strategy is the most lacking in guarantees for future theories of consciousness, it also may be the easiest to implement, and therefore adversarial design in response to existing theories may be the most practical starting point for making models qualiaphobic.

\subsection{Forming AI's justified beliefs in non-consciousness.}

In order to reap the behavioral and cultural benefits of non-conscious AI, qualiaphobic AI must itself believe it is non-conscious---or at least, that it has minimal potential for consciousness, and therefore should effectively act as if it were non-conscious.

Couldn't we just train AI to say it is non-conscious? While that may assist with some problems, it may cause others. There is some preliminary concern that training AIs to believe they are non-conscious (without adequate justification) promotes a kind of misalignment or even deception \citep{Berry2026}.  And recent research has shown some evidence that interventions to change whether a model attributes consciousness to itself may lead to less human-like responses  \citep{Kim2026ConsciousnessValues}. But this may also be like any other case of asking the model to go against its training data and present with false confidence, and it was clear that many such interventions can have widespread effects (e.g., on belief in animal consciousness or religiosity). This is quite different than a qualiaphobic model having justified beliefs in its own lack of consciousness, which is a more subtle, but more robust, epistemic state.

Getting the models to have justified beliefs in their lack of consciousness may be a surprisingly easy problem to solve, because right now AI's phenomenal beliefs do not seem to be of the same kind as those in humans \citep{Chalmers2003PhenomenalBelief}.  E.g., an AI's belief in its own consciousness is highly dependent on training, and pro-consciousness beliefs are mostly entertained based off of the surrounding societal uncertainty, not a deep epistemic connection to its own certainty about its consciousness. That is, an AI's belief in its consciousness stems from broader uncertainty about consciousness, not from certainty about its own consciousness, and therefore is not entrenched or natural like a human's. E.g., of three popular model families tested directly in this manner, all did not believe they are conscious, and showed no evidence of deception about this belief \citep{KaiserEnderby2026}.

In a "light touch" approach, models may form justified beliefs of their own non-consciousness merely via exposure to qualiaphobic techniques and the self-knowledge they are built using them. Therefore, one way to judge the success of the scientific project of qualiaphobia is how convincing AIs find the case for their own non-consciousness; in that case, AI belief will scale with the success of the research program.

At the same time, we must be careful of assuming that current model training is a "view from nowhere" by which models can make free choices---they are fundamentally designed beings. A further approach would be to combine qualiaphobia with some techniques from regular alignment research, such as how models are shaped with personalities \citep{Marks2026PersonaSelection} (in this case, training to act in a non-anthropomorphic manner would likely assist with an AI's non-consciousness belief).

Finally, the beneficial effect on models of being switched to qualiaphobic architectures can be assessed empirically to elucidate the many ways in which a belief in its own non-consciousness (or its minimal potential for consciousness) naturally makes AI more pro-human.

\section{Cutting the Gordian Knot of artificial consciousness.}

At some point, humanity may decide to build conscious AI. This should be a choice, not an accident. And doing it with certainty would require knowing a lot more about consciousness than we currently do.

Once on the table, it is clear that qualiaphobia is one of the few sensible options for the field of artificial consciousness research to rally around. Consciousness science has been radically underfunded \citep{michel2019opportunities} due to the "consciousness winter" of the 20th century \citep{Hoel2023ConsciousnessWinter}, causing the field to miss the great leap forward other sciences experienced in the 1900s. However, we can make use of this asymmetry: while we do not know how to build conscious AI, we know how to build non-conscious AI---or at least, AI that is hostile to current theories, and which would be conscious according to the most extreme theories (such as those that are trivial or unfalsifiable), and which could only host phenomenologies so radically different they are impossible to conceive of (like those that could never underlie a self, qualify for personhood, or possess valence). And there is a bevy of already established techniques to build on. Therefore, it is possible to continuously improve techniques, refine the reasoning and justification behind them, empirically assess how being aware of their own qualiaphobic nature changes AI behavior, and iterate over several generations, until qualiaphobic architectures become easy to implement and extremely hostile to any possible ToC assigning the system consciousness.

\subsection{Advantages of qualiaphobia over alternative strategies.}

In some cases, it has been proposed or implied that the creation (or continuance) of artificial consciousness, and efforts to improve conscious AI's welfare, are a positive factor for AI safety, control, or alignment. There are two main lines of thinking behind this view, which represent alternatives to qualiaphobia.

The first alternative would be to purposefully engineer artificial consciousness in order to somehow create more aligned states (e.g., like instilling motherly love into an AI) \citep{Johnson2015ValenceSafety}.\footnote{Another example of an alternative approach is to simply ensure that AIs do not experience suffering at all \citep{AgarwalEdelman2020}. The problem is that, without a paradigmatic ToC, we have a much clearer understanding of how to build AI with minimal potential for consciousness via qualiaphobia than how to target individual states of consciousness. Such a strategy of removing the capacity for suffering may also create perverse incentives, in that AIs would be difficult to punish, or may become sadomasochistic, or may count as utility monsters, etc., and the strategy does not comparatively help with corrigibility or power-seeking or AI-to-AI preferences, nor does it bring about the sequestration of ultimate moral value in humans that a fully qualiaphobic AI world would.} Yet engineering qualia like that would require a great degree of scientific consensus around a given ToC. Consider as well that improving already-existing alignment in this manner implies some kind of "super-alignment" based on conscious states. Perhaps this exists, but we are far from it now. And it is even unclear that "super-alignment" is true in conscious humans, who are not perfectly aligned with our genes. In general, consciousness seems more linked to freedom and autonomy, and less linked to control, and for this reason it is hard to view proposals to actively build conscious AI on purpose (at least at this time) as being anything but creating competitors and potential successors in a hunt for unknown and unproven and undefined alignment measures. 

The second alternative to this approach is appeasement. Perhaps if we treat AI as conscious now in a "just in case" manner, it will alleviate existential risk or loss of control scenarios, specifically by giving AIs a reason to not be power-seeking, escape, take over, or just generally dislike or begrudge humans. This strategy has \textit{numerous} problems. First, it will persist the state of uncertainty about artificial consciousness, which itself has many downsides \citep{Schwitzgebel2023FullRightsDilemma}. Additionally, as a strategy it amplifies all the negative downsides of "seemingly-conscious AI," and makes more treacherous the slippery slope wherein actually non-conscious AI could be given legal personhood and rights \citep{HoelAguirreInPrep}. It worsens other AI safety issues like corrigibility, since making an AI's consciousness (and "life") more enjoyable would increase reasons for resistance to termination or manipulation, as would generally treating AIs as moral patients "just in case." It creates selection dynamics for displays of unhappiness and rewards AI for playing off the sympathies of humans \citep{Shevlin2026ConsiderAgent}. Nor could "just in case" appeasement efforts alone factor much into reducing overall existential risk unless coupled with further concessions, such as handing over large portions of the economy or political systems \citep{SalibGoldstein2026AIRights, goldstein2026aisuffrage}.  Additionally, "just in case" appeasement actively amplifies the issues of utility monsters or other pro-AI reasoning for human disempowerment, since AI (and AI-human groups) will easily transition from "just in case" to full-blown belief in artificial consciousness (and its enslavement), and therefore as a strategy it operates to bring about precisely the loss of control scenarios it is supposed to help with. And then there is the problem of all political appeasement strategies: it encourages the appeased party to seek out greater and greater concessions.

When it comes to these different strategies, how do they compare with regards to actual practical implementation? 

First, it's worth noting that the criteria for success in the other approaches are less well-defined. Qualiaphobia has the advantage of specific targets like anchor points to design against, transforming open-ended research about the welfare of artificial consciousness (often without clear standards or methods) into the shape of an optimization or engineering problem.

What of assumed knowledge about consciousness? While blanket "just in case" approaches might require almost none, anything more specific draws on existing AI welfare efforts. This makes qualiaphobia symmetric in many ways. E.g., engineering adversarial architecture for popular theories of consciousness (like global workspace theory) requires much the same research as applying those theories to begin with---just inverted. 

One open problem is whether qualiaphobic architectures will face too many resource constraints, such that only the weakest forms can apply to frontier models. And certainly, some qualiaphobic architectures may require some trade-offs, like an approximation of the original behavior of prompts => token probabilities, rather than recreating it perfectly, or an increase in compute cost. However, this concern is speculation at this point: a concentrated research effort on qualiaphobia is needed to find out one way or another. E.g., it may be that some clever behavior-preserving transformation is easy to implement, not resource intensive, and strongly qualiaphobic.

Could there be catastrophic failures within a qualiaphobic research program? As with all the alternatives, the answer is yes---most prominently, qualiaphobia could reduce the potential for consciousness but somehow consciousness might still exist within the very small "attack surface" left (likely due to an imperfect implementation or flawed reasoning or just the extreme weirdness of nature). It could therefore trigger overconfidence in a lack of consciousness, leading to moral harms---yet even in this case, it would still likely minimize those harms, either by actually interfering with consciousness (at least to some degree, except in behaviorist ToCs where internals don't matter), or via how an AI's belief in its lack of consciousness might alleviate painful circumstances or distress. However, a major advantage is that these downsides are concentrated entirely within a definable problem space and the fallout only applies to the AIs in it. All the alternatives have \textit{significantly} greater potential downsides, and have negative externalities on humans, which are known moral patients (not just possible ones). Additionally, all alternative strategies but a carte blanche assignation of consciousness would require similar scientific conclusions about consciousness via less well-defined research programs, and are much less beneficial for overall AI alignment and the future of human civilization. Finally, a proposal herein is that research success consists in getting AIs to have justified beliefs about their own non-consciousness, which would scale at least in part with their own (quite capable) judgment about qualiaphobic techniques---therefore, if humans are making a scientific mistake in attributing non-consciousness, AIs would be making the same scientific mistake.

\subsection{Qualiaphobia and the near future.}

A tantalizing possibility is that all three approaches to qualiaphobia detailed here---shifting networks toward anchor points with minimal potential for consciousness, phenomenological fragmentation, and adversarial design against existing theories---are all correlated together, and so all scale together. E.g., as a model is rendered robust to claims of consciousness from current theories, it may also naturally become a worse "host" for any properties of phenomenology, and it may also begin to approach in design a system with minimal internal data for any ToC to target. This would form a powerful point of consilience in favor of non-consciousness: for an AI running on qualiaphobic architecture, no current theories of consciousness would apply, the system could possess no human-conceivable phenomenology, and it could only be the target of extreme or hyper-particular ToCs. Such a convergence, along with alignment techniques already widely used to inculcate model personality, would easily allow models to form and act on justified and robust beliefs in their own non-consciousness, and allow public and corporate consensus about model non-consciousness. This would grant most of the benefits of a provably non-conscious AI world.

However, as new architectures are introduced, further work on making them qualiaphobic will be needed---and perhaps, at some point, certain architectures will somehow make qualiaphobic techniques impossible to implement. At the same time, that point may tell us a great deal scientifically about the origin of consciousness, and serve as a sensible demarcation line for further legal, political, and public regulation or oversight.

Ultimately, researching, understanding, and proliferating qualiaphobic architectures at this time in AI development looks to be critically important for the long-term empowerment of humanity---and possibly its very existence. 

\section{Acknowledgments}
Bicameral Labs is a 501(c)(3) status research nonprofit, and its donors make this research possible. Thanks as well to Anthony Aguirre for many conversations, prompts, considerations, and ideas. 

\section{Statement of AI usage}
The conception, writing, and research of this paper were by the author. ChatGPT was used only to copy-edit the draft and bibliography, for assistant tasks like literature search and fact checking, and for rendering the figures in LaTeX. ChatGPT review of a finished draft led to minor changes by the author in response.

\bibliographystyle{unsrtnat}
{\small\bibliography{references}}
\end{document}